\documentclass[sigconf,10pt]{acmart}
\usepackage[utf8]{inputenc}
\usepackage{graphicx} 
\usepackage{xspace}
\usepackage{gensymb}
\usepackage{mathtools}
\usepackage{multirow}

\usepackage{amsfonts}

\usepackage{caption}

\def\block(#1,#2)#3{\multicolumn{#2}{c}{\multirow{#1}{*}{$ #3 $}}}

\newtheorem{definition}{Definition}
\usepackage{amsthm}
\newcommand{\name}{\textit{GEM}\xspace}
\newcommand{\transfer}{\textbf{Transfer}\xspace}
\newcommand{\swap}{\textbf{Swap}\xspace}

\setcopyright{none}
\author{Shubham Rohal}
\email{srohal@ucmerced.edu}
\affiliation{%
  \institution{UC Merced}
  \country{USA}
}

\author{Dong Yoon Lee}
\email{dlee267@ucmerced.edu}
\affiliation{%
  \institution{UC Merced}
  \country{USA}
}

\author{Phuc Nguyen}
\email{vp.nguyen@cs.umass.edu}
\affiliation{%
  \institution{UMass Amherst}
  \country{USA}
}

\author{Shijia Pan}
\email{span24@ucmerced.edu}
\affiliation{%
  \institution{UC Merced}
  \country{USA}
}

\title{\name: Gear-based Environment-Integrated Mobility for Adaptive Indoor Human Sensing}

\begin{document}

\begin{abstract}
Infrastructure-based sensing systems, like Wi-Fi-, thermal-, vibration-based approaches, provide continuous and unobtrusive indoor human monitoring services.
They are often deployed statically for long-term continuous monitoring, which often leads to inefficient sensing/inflexible deployment due to human mobility or high maintenance/data volume for dense deployments.
In contrast, autonomous and human-carried mobile devices can better adapt to human mobility. 
However, their physical presence (e.g., drones or robots) may induce observer effects, while their operation often imposes additional burdens, such as wearing (e.g., wearables) and frequent charging. 

We present \name, a hybrid scheme that introduces the mobility to infrastructure-based sensing.
\name integrates a matrix of gears into everyday surfaces (e.g., floors, walls) to turn them into ``public transportation" for moving infrastructure sensors around.
We design and fabricate a 3 $\times$ 3 gear matrix prototype, that can effectively move sensors from one location to another.
We further validate the scalability of the design through simulation of up to  $64 \times 64$ gear matrix with concurrent sensors.

\end{abstract}
\begin{teaserfigure}
  \includegraphics[width=\linewidth]{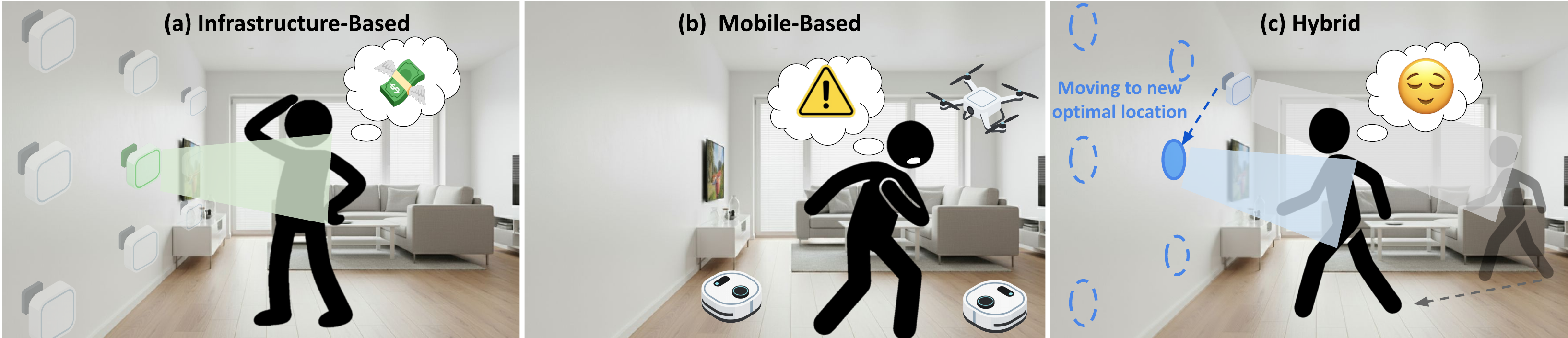}
    \caption{\name design intuition. (a) depicts the infrastructure-based sensing system relies on dense deployment to achieve high fidelity sensing, which is costly. (b) shows the mobile-based sensing system enables effective sensing via mobility. However, they may interfere with natural human behavior and are constrained by  charging requirements. (c) introduces a hybrid design, where traditional infrastructure-based sensing system can adapt their deployment location for high fidelity signal collection while remain unobtrusive.}
  \label{fig:teaser}
\end{teaserfigure}

\maketitle

\section{Introduction}
In this paper, we show how a sensor system can be both environmentally integrated and mobile.
Sensing systems generally fall in two categories, infrastructure-based and mobile-based systems.
They have complementary properties \cite{pan2020opportunities} in terms of spatiotemporal coverage and power availability.
For example, infrastructure-based (ubiquitous) systems, such as Wi-Fi-\cite{wu2017device}, vibration-\cite{pan2019fine}, thermal-\cite{zeng2025thermikit}
, offers long-term, continuous, and unobtrusive human sensing solutions for indoor environments.
However, these systems rely on static sensors and often suffer from inefficiencies in sensing and deployment, high maintenance overhead, and large data volumes from dense installations.
On the other hand, mobile platforms, such as autonomous (drones and indoor robots) and on-body mobile systems (smartphones and wearables), provide an alternative solution that mitigates these issues by enabling fewer sensors to efficiently collect data across a large area.
For example, drone-based mobile platforms, such as aerial platforms \cite{mao2017indoor} and ground-based platforms \cite{eslava2024re}, reduce the number of sensors needed to cover the sensing area by following the monitoring target.
Human-carried devices, such as earable \cite{hossain2019human}, intrinsically follow the target by being worn on the body.
While these solutions enable a sparse deployment to sense users over a greater area, they either interfere with natural human behaviors due to their physical presence and the user’s awareness of being followed or impose the burden of continuous wearing and charging.
Therefore, this paper aims to answer the question: \textit{How can we move infrastructure sensors discretely to monitor people?}

We present \name, a \underline{G}ear-based \underline{E}nvironment-integrated \underline{M}obility platform for indoor sensors.  
The intuition is that, unlike traditional mobile systems that use individual actuators for each device, \name creates a mobility layer between the environment and sensing devices, providing a ``public transportation system'' for sensor ``passengers''.
Just like a railway system, \name uses ``rails'' to enable sensor movement.
However, the rail-based mechanism restricts motion to a linear path. 
Therefore, we investigated other types of motion -- rotation -- with meshing gears to transmit motion. 
\name integrates gear and rail design by aligning a matrix of gears to enable synchronized rotation and cross-gear movements.
However, shared resources may result in collisions.
We develop a multi-sensor path finding algorithm for physically constrained movements using \name, and provide a proof that for a $n \times m$ gear matrix, the maximum number of sensors that can be moved to any target location is $\lceil{\frac{n \times n }{2}}\rceil$.
In summary, this paper makes the following contributions:
\begin{itemize}
    \item We present \name, a hybrid design that enables mobility for infrastructure-based sensing systems. 
    \item We design the hardware of a sensor movement platform that integrates gear- and rail-based mechanisms to achieve flexible motion across ambient surfaces.
    \item We develop a path-planning algorithm with collision avoidance for the gear-based mobile platform, grounded in a formal proof of physical movement constraints.
    \item We demonstrate the system's feasibility with a $3 \times 3$ real-world prototype, and illustrate scalability via simulations under varying sensor deployment densities.
\end{itemize}
\begin{figure}[!t]
    \centering
    \includegraphics[width=0.9\linewidth]{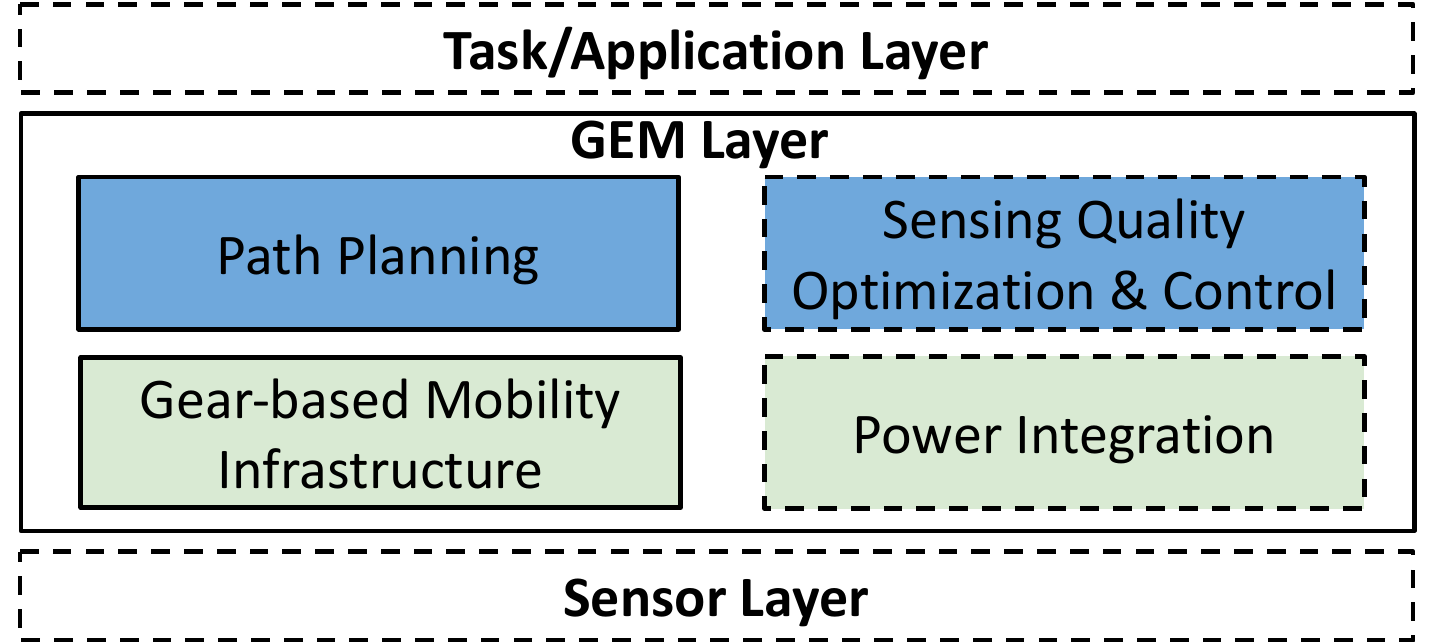}
    \caption{\name overview. We consider \name provides a layer of services to enable mobility of an infrastructure-based sensing system. The solid-line boxes indicate the scope of this paper, while the dashed-line boxes represent other aspects or future work. Green boxes are hardware designs, and blue boxes are software designs.}
    \label{fig:layer}
    \vspace{-5pt}
\end{figure}
\section{Proposed Approach}
We consider \name a modularized layer in the infrastructure-based sensing systems, providing mobility services. Figure \ref{fig:layer} depicts the position of the \name platform, where it interfaces with the task/application layer and sensor layer.
This paper focuses on the enabling of mobility with the infrastructure hardware (Section \ref{sec:sensor_node}) and path planning software (Section \ref{sec:path-planning}).
In the future, we will further explore (1)
the sensing data and task/application requirements will be integrated at \name layer for optimal sensor placement, and
(2) the integration of the power and data transmission hardware to the infrastructure hardware introduced here.

\subsection{Gear Matrix Enabled Mobility Design}\label{sec:sensor_node}
We use gears as the assistive structure for two key reasons: (1) they enable \textbf{fine-grained} directional control via rotation, supporting repeatable spatial calibration and controlled sensor movement along surfaces; and (2) they provide \textbf{scalable} coordination by transferring motion across connected components.
The intrinsic mechanical coupling of gears enforces synchronized motion among connected components. 
This synchronization allows multiple actuation points to move coherently, which is a key requirement for aligning sensing modules that operate across adjacent rotational segments. 
The interlocking teeth of the gears yield deterministic angular displacement to enable precise modeling and control of motion. 
This design supports repeatable spatial calibration and controlled sensor movement along curved or rotating surfaces. 
Last but not least, gears maintain alignment through minimal slip and stable torque transfer, which provides a robust foundation for adaptive surface sensor reconfiguration.

\begin{figure}[!t]
    \centering
    \includegraphics[width=0.6\linewidth]{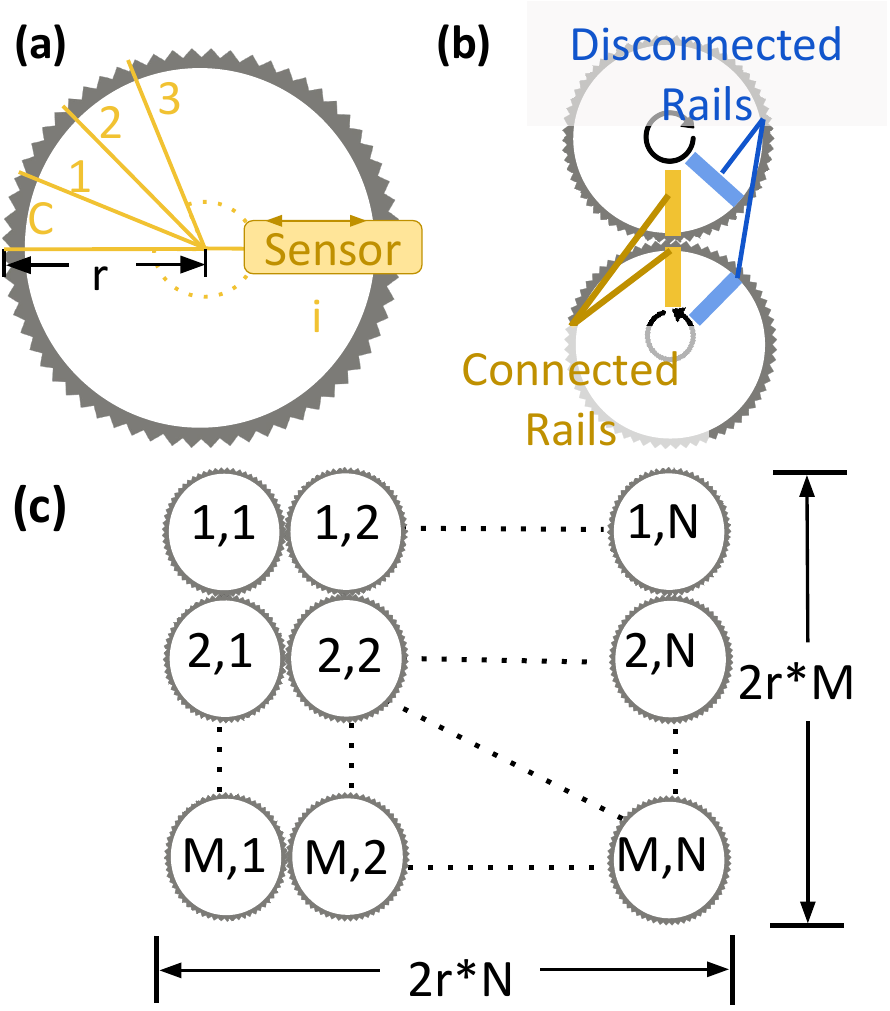}
    \vspace{-1ex}
    \caption{\name design overview. (a) individual gear design with radius of $r$ and channel number of $C$. (b) rail sensor transfer mechanism. (c) gear matrix of $M \times N$.}
    \label{fig:overview}
\end{figure}

Figure \ref{fig:overview} depicts \name design for infrastructure sensor mobility, including (a) an individual gear structure design for sensor mounting, (b) a pairwise gear configuration for sensor transfer, and (c) a gear matrix enabling multi-sensor 2D movement.
First, we design the sensor mounting mechanism to be compatible with the fine-grained directional control.
As shown in Figure \ref{fig:overview} (a), each gear structure has $C$ channels evenly distributed across its surface marked in yellow lines.
Each channel has a rail structure, where a sensor can be mounted on it and move along it, as shown on the channel $i$.

For two neighboring gears, we align the rails on them as shown in Figure\ref{fig:overview} (b).
When they rotated to the designated angles, the corresponding rails will connect, as marked in yellow lines, allowing the mounted sensor on one gear transfer to the other one.
For the disconnected rails, as marked in blue lines, we expect there is at most one sensor on such pairwise rails. Otherwise, when the gears rotate to the angle that aligns them, the sensors will not be able to transfer to the connecting gear. 
We refer to this scenario as collision, which we will introduce in later sections as the ``\textit{Gear Channel Parity}" problem.


\name uses a matrix of gears of same size on the target surface, as shown in Figure \ref{fig:overview} (c).
For the gear at location $(i,j)$ in the matrix, its channels are aligned with those of its neighboring gears, i.e., $(i\pm1,j\pm1)$.
Only one gear in the matrix is actuated, and all the gears in the matrix will have synchronized rotation.
For each pairwise neighboring gears, their rotation direction are opposite.
For example, if $(i,j)$ rotates clockwise, $(i\pm1,j\pm1)$ rotate counter clockwise.

\subsection{Constrained Path Planning} \label{sec:path-planning}
When multiple sensors move simultaneously on the gear matrix, they might not be able to transfer to the neighboring gear if the rail is already occupied.
We call this type of collusion as the \textit{Gear Channel Parity}.
To allow multiple sensors to move along the gear platform without collisions, we design an algorithm based on a formal proof.
%
\subsubsection{Proof of Collision-Free Capacity under Gear Channel Parity Constraints} 
We first derive a formal proof to model how multi-agent path finding work with movement constrained defined by gear matrix.
For the proof of concept, we consider a simplified case of a $n \times n$ matrix of $c$ channel gears.
We prove that the maximum number of sensors that can move to any target location collision-free is $\lceil{\frac{n \times n }{2}}\rceil \times c$. 
We first define a base case for the smallest configurations of a single-channel gear platform, $G_2$.
Next, we define different ways we can move sensors across the platform with the \transfer and \swap operations. 
Last, we use an inductive proof to show the feasibility of larger gear configurations $(n+1) \times (n+1$), given that a smaller configuration of $n \times n$ is possible.


\begin{definition}[$G_2$]
\label{definition:g2}
Let $G_2$ be a set of states for gear matrices with 2 sensors, $s_0$ and $s_1$, placed in this pattern: \[
    G_2 = \left\{\\
    \begin{bmatrix}
    s_0 & \_ \\
     \_ & s_1\\
    \end{bmatrix},%
    \begin{bmatrix}
    \_ & s_0\\
     s_1 & \_ \\
    \end{bmatrix},%
    \begin{bmatrix}
    \_& s_1\\
     s_0 & \_ \\
    \end{bmatrix},%
    \begin{bmatrix}
    s_1 & \_ \\
     \_ & s_0 \\
    \end{bmatrix}\right\}
\]

For every element $i$ and $j$ in G2, there exists a $\mathbf{\mathbf{Transfer}_{i\to j}}$, where a set of operations can transform from $i$ state into $j$ state.
We formalize this definition as such:
\[
    i,j\in G_2 \iff \mathbf{Transfer}_{i\to j}(i) = j
\]
\end{definition}

\begin{definition}[checkerboard pattern]
A matrix is said to have a \textbf{checkerboard pattern} if all $2 \times 2$ submatrices of a matrix are an element of $G_2$.
For any 2 matrices, $g_{\mathrm{1}}$ and $g_{\mathrm{2}}$, that both have \textbf{checkerboard patterns} and the same shape $X \times Y$, 
they are transferable to each other, if every $2 \times 2$ submatrix in $g_\mathrm{1}$ maps to every $2 \times 2$ submatrix in $g_\mathrm{2}$ at the same location, when \textbf{Transfer} operation is applied.
We model the mapping of two checkerboard with following premise:
\[
\begin{split}
     \forall x \in (0,\dots,X-2)&,\forall y \in (0,\dots,Y-2) \\
    (T(g_\mathrm{1}[x:x+2,y:y+2]&, g_\mathrm{2}[x:x+2,y:y+2])) \\
    \implies \mathbf{Transfer}_{g_\mathrm{1}\to g_\mathrm{2}}(g_\mathrm{1}) &= g_\mathrm{2}
\end{split}
\]

The \textbf{transferrability} of matrices that have \textbf{checkerboard patterns} are also transitive, such that: \[
    \forall g_i,g_j,g_k \mid T(g_i,g_k) \land T(g_k,g_j) \implies T(g_i,g_j)
\]
\end{definition}

\begin{proof}[In-Place Swap]
\label{lemma:swap}
To allow the each pathfinding sensor to not interfere with other pathfinding sensors, we require a mechanism to swap two nearby sensors in place, such that they do not collide nor they affect other sensor locations.
For this in-place swap proof, we show how we can use several \transfer operations on an initial state matrix $g_1$ to achieve a target state $g_2$ without changing any other sensors in the matrix other than the two sensors $s_0$ and $s_2$.

For any $3 \times 2$ submatrix $g$ of an $n \times n$ matrix that follows the gear matrix checkerboard pattern, we can perform an in-place swap operation \swap on a matrix $g_1$, such that we get the resulting $g_2$ matrix: \[ 
\begin{split}
    g_\mathrm{1} = \begin{bmatrix}
    s_0 & \_ & s_2\\
     \_ & s_1 & \_ \\
    \end{bmatrix},
    g_\mathrm{2} = \mathbf{Swap}(g_\mathrm{1}) = \begin{bmatrix}
    s_2 & \_ & s_0\\
    \_ & s_1 & \_ \\
    \end{bmatrix}
\end{split}
\]
Due to the transitive property of the \textbf{Transfer} operation, we prove that $g_\textrm{1}$ can transform to $g_\textrm{2}$. Therefore, we define the \textbf{Swap} operation as an element of all possible \textbf{Transfer} operations as such: 
    $\mathbf{Swap}(g_\textrm{2}) = (\mathbf{Transfer}_{1,2} \circ \mathbf{Transfer}_{0,3} \circ \mathbf{Transfer}_{1,2})(g_\mathrm{1})$
\end{proof}

\begin{proof}
For any $n \times n$, $\forall n \in \mathbb{N}^+$, $n \geq 2$ gear matrix, let there be a set of sensor 
location states $G_n$, each with size $n \times n$, such that each element in $G_n$ is a matrix containing $\lceil{\frac{n \times n }{2}}\rceil$ sensors in a \textbf{checkboard pattern}, such that each gear matrix has a possible path to any other gear matrix in $G_n$. 
We wish to prove that the successor, $G_{n+1}$, is a set in which all states, each with size $(n + 1) \times (n + 1)$, have a possible path to any other element in $G_{n+1}$.
Suppose we model each element in $G_{n+1}$ as some matrix where $\forall g_{n+1} \in G_{n+1}$, such that when we make several arbitrary \textbf{Swap} operations, there exists a state in $g_n$ that is equivalent to a submatrix in $g_{n+1}$, $\exists g_{n} \in G_{n} \mid g_{n+1}[0:n][0:n] = g_{n}$:
\[
    g_{n+1} = \begin{bmatrix}
        s_1 & \_ & s_2 & \_ &...& & s_{\lceil\frac{n + 1}{2}\rceil} \\
        & & & & & & \_ \\
        \block(4,4){g_{n}} & & & s_{\lceil\frac{n + 1}{2}\rceil + 1} \\
        & & & & & & \_ \\
        & & & & & & ... \\
        & & & & & & s_{n + 1}
        
    \end{bmatrix}
\]
Since $g_n$ is \textbf{transferrable}, all $2 \times 2$ submatrices in $g_{n+1}[0:n][0:n]$ are also \textbf{transferrable}.
For the ordered sequence of the rest of the sensors not in $g_n$ $s: \forall s \notin g_{n}$, any two neighbors can be swap in-place with proof ``In-Place Swap''.
Therefore, the sequence can be rearranged in any order with in-place swapping only.
In this way any state $g_{n+1}$  can swap with other states in $G_{n+1}$.
Since matrix $g_{n}$ and sequence $s$ are both \textbf{transferrable}, all $2 \times 2$ submatrices from $g_n \cup s$ are \textbf{transferrable}.
Since both $g_n$ and $s$ cover $g_{n+1} = g_n \cup s$, according to the definition of \textbf{transferrability}, $g_{n+1}$ is \textbf{transferrable}.
\end{proof}

\subsubsection{Path Planning with Parity Constraints}
For gear-based path planning \name modifies the traditional $A\ast$ path finding algorithm \cite{BibEntry2021}.
We can model our problem such that for any intermediate state $g_n$ during the path-finding process, \textit{\textbf{p(n)}} is the cost to reach the current state from the initial state and \textit{\textbf{h(n)}} is the estimated cost to reach the target state.
We then try to minimize the total cost function \textit{\textbf{f(n) = p(n) + h(n)}}.
To account for the total cost of a path, we consider \textbf{r($\theta$)} as the rotation cost and \textbf{a} as the cost per sensor to move across a connected rail.
If the total \textit{\textbf{q}} sensors are moving for a state change, the cost of the current state change can be defined as \textit{\textbf{p(n) = r($\theta$) +(q $\times$ a)}}.
To calculate the heuristic cost \textit{\textbf{h(n)}}, we aggregate the Manhattan distance from the current to the target location for all sensors.
The path-finding algorithm then uses the \textbf{Transfer} and \textbf{Swap} operations mentioned in Section \ref{sec:path-planning} to find possible paths.
To take \textit{Gear Channel Parity} constraints into account, we incorporate collision filtering to remove any path that moves two sensors to the same connected channel.
We weigh the remaining paths with the cost function \textit{\textbf{f(n)}} as priority, to find the heuristically shortest path.
The synchronous gear rotation also allows multiple sensors to move concurrently,
and this simultaneous rotation is counted as one simulation step.
We designed our path-planning to consider the movement of all sensors and decrease the total path steps to reach the target state.
We also design a dynamic masking technique to track the connected channel pair across the gear matrix for each rotation step. 
This masked connection matrix is applied to the current sensor's location to update all possible paths.

\section{Prototype and Evaluation}
In this section, we demonstrate the feasibility of the proposed approach through a real-world prototype implementation and characterization. 
The correctness of the constrained path planning concept is validated via scale and population analysis with simulation.

\subsection{Prototype and Characterization}
\begin{figure}[!t]
    \centering
    \includegraphics[width=0.85\linewidth]{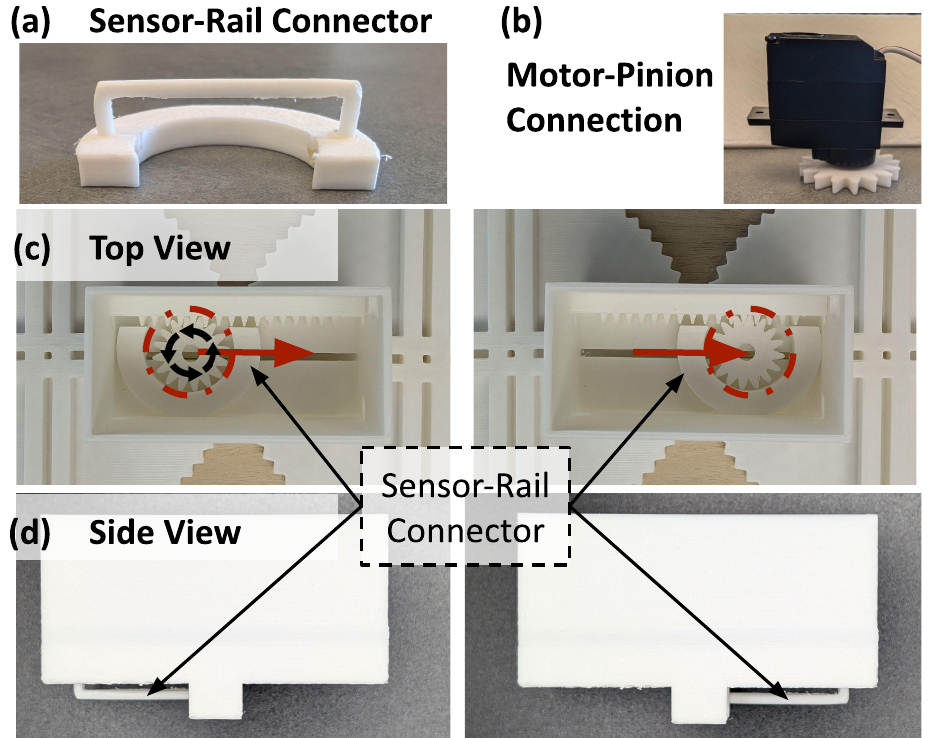}
    \caption{\name implementation with 4 channels.}
    \label{fig:sensor_node}
\end{figure}

\begin{figure}
    \centering
    \includegraphics[width=\linewidth]{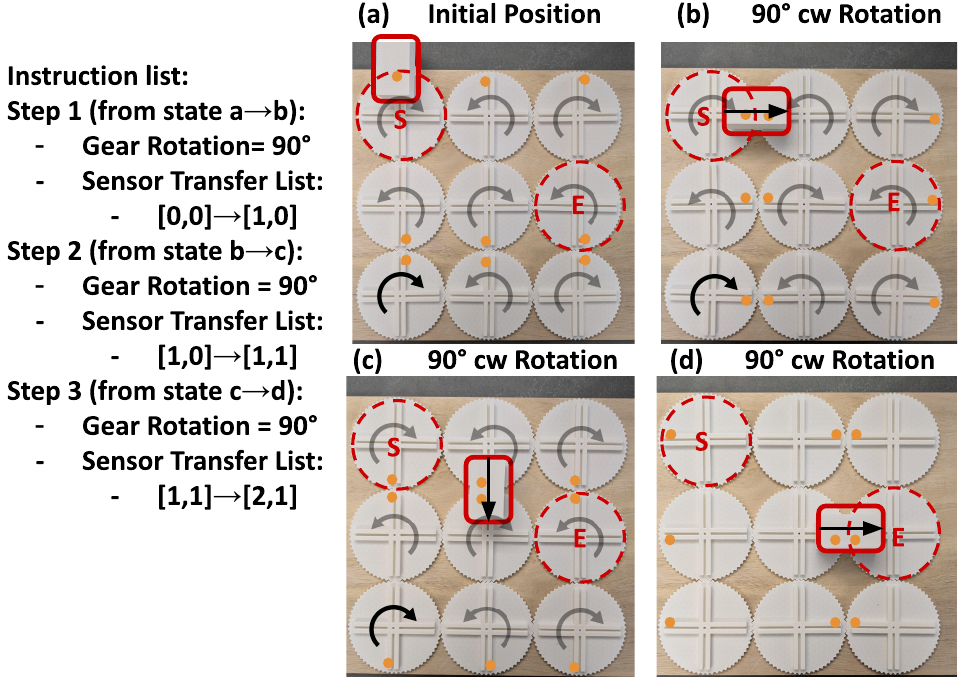}
    \vspace{-2ex}
    \caption{Example path finding (from $S$ to $E$) and sensor/gear movement.}
    \label{fig:parity}
\end{figure}

We implement a real-world working prototype with $3\times3$ gear matrix of 4 channels, as shown in Figure \ref{fig:sensor_node} and \ref{fig:parity}. 
We depict the connector that mounts the sensor to the rail in Figure \ref{fig:sensor_node} (a) and the servo motor used as the actuator in Figure \ref{fig:sensor_node} (b).
We implement the inter-gear transfer with a servo motor connected to a small gear pinion in the \textit{rack and pinion} mechanism \cite{laithwaite1970rack}, as show in Figure \ref{fig:sensor_node} (c) and (d).
When the servo motor rotates, the circular motion converts into linear lateral motion, which moves the sensor-rail connector along the rail, as marked in red arrows.

In order to coordinate gears and multiple sensors for path finding, we implement a structured interface between the path-planning algorithm and hardware.
Figure \ref{fig:parity} gives an example of how the hardware follows a list of steps that both rotate the gears, and transfer each sensor to different gear channels.
For each step, we first rotate the gear such that the channels of moving sensors align with other nearby gear channels, and then actuate sensors in the transfer list to the new gear.

The prototype's movements consist of two actions: rotating the gear and transferring the sensor between adjacent gears.
We characterize the time for these two actions by repeating the action 10 times. 
The rotating gear by 90$\degree$ takes 0.38 ($\pm$0.025) seconds. 
The sensor movement across connected channels takes 0.56 ($\pm$0.023) seconds.

\subsection{Gear Matrix Size Analysis}
\begin{figure}
    \centering
    \vspace{-2ex}
    \includegraphics[width=0.9\linewidth]{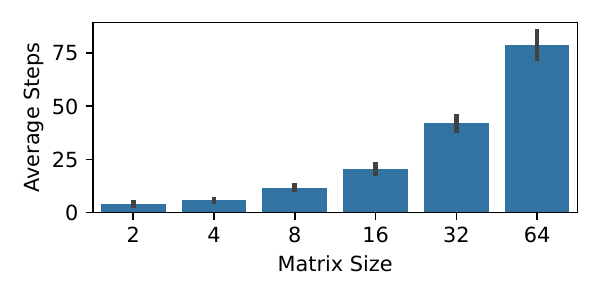}
    \vspace{-2ex}
    \caption{The change in average steps required for sensor relocation with increase in gear matrix size $n\times n$}
    \label{fig:grid_size}
\end{figure}

To understand the system capability for scaling up in the size of the gear matrix,
we conducted a simulation experiment with 4 sensors and the matrix size of $n \times n$, where $n$ ranges from $2$ to $64$ exponentially, and report the average steps took to reach the target location.
We randomly select the start and target locations for all four sensors and repeat the experiment 16 times for each gear matrix size.
Figure \ref{fig:grid_size} shows the average steps it takes for 4 sensors to reach the target location for each matrix size.
We can observe that the increase in average steps is proportional to the increase in matrix size,
which indicates the effectiveness of the path-finding algorithm, which finds near-optimal solutions close to the Manhattan distance.

\begin{figure}
    \centering
    \vspace{-2ex}
    \includegraphics[width=0.9\linewidth]{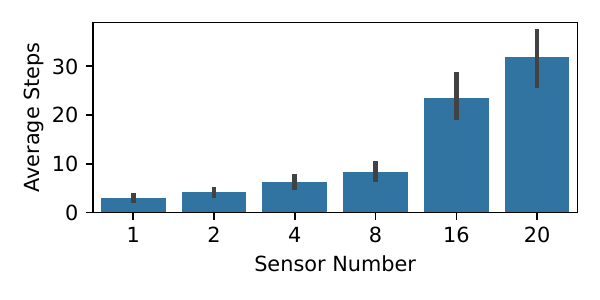}
    \vspace{-2ex}
    \caption{The average steps required for sensor relocation with an increase in the number of sensors on gear matrix of size $3 \times 3$.}
    \label{fig:sensor_pop}
\end{figure}

\subsection{Sensor Population Analysis}
The sensor population is another factor that impacts the path-finding capabilities of \name framework.
To evaluate this, we simulate a gear matrix of $3 \times 3$ with 4 channels.
We changed the number of sensors \textit{\textbf{q}} by the power of 2 until we reach the maximum capacity (20 sensors in this case) as discussed in Section \ref{sec:path-planning}. 
We randomly selected the start and target locations for each sensor and recorded the steps the sensor took to reach the target location.
We repeated the experiment 16 times for different sensor populations,
Figure \ref{fig:sensor_pop} shows the average steps it takes for a sensor to reach the target location.
We can observe that with an increase in sensor population, the average steps required also increase.
This is because, as the number of sensors increases, the total available channel space on different gears decreases, resulting in sensors taking a longer path. 
We also observe that the standard deviation among required steps increases as the number of sensors increases.

\section{Next Steps}
Although the preliminary results are encouraging, multiple research questions remain unanswered, leaving space for future research activities. 

\paragraph{Path-Finding Optimization}
In this work, we implement a greedy path-finding algorithm for a multi-agent path-planning scenario, which is not ideal.
Some cases exist in which the current system can run into a bottleneck state.
In the future, we want to optimize this to minimize the required steps.
Building upon rich literature on multi-agent path planning in the field of robotics \cite{wang2020multi,hu2023anti}, we will explore the surface structural constraint-based algorithm adaptation.

\paragraph{Power Integration}
In the current prototype, the sensor is powered through a wired connection. 
However, in the presence of multiple sensors, the power wires can get entangled with each other.
To address this issue in the future, we plan to explore gear design with an embedded power connection for the sensor.
The current gear design uses a rail track to connect sensors to the gear and inter-gear transfer.
Embedding a power circuit in rail tracks could be a potential solution.

\paragraph{Sensing Quality Optimization}
In this work, we assume the optimal/ end location of the sensor is already known to the path-finding algorithm.
Optimal sensing location estimation can be a challenging task on its own.
In the future, we plan to include a sensing data quality optimization layer that can predict the optimal sensing location dynamically.


\section{Related work} 
This section discusses the related work for sensor mobility via autonomous movement systems.
The sensing systems are integrated in autonomous platforms such as ground, water robots, and aerial drones to dynamically change sensor location for numerous applications, such as agriculture \cite{van2018autonomous, bai2023vision}, water quality monitoring\cite{kazeminasab2020development}, surveillance, and remote delivery\cite{fahlstrom2022introduction, leonard2016autonomous}, sea exploration, and rescue\cite{sahoo2019advancements} or rail based mobile platforms for pipeline surveillance and maintenance\cite{liu2023autonomous, kahnamouei2023comprehensive}.
Although these methods are highly effective for outdoor applications, their adaptation to indoor applications presents its own challenges, like sensor error accumulation \cite{Liu2024Feb} and high computational overhead.
Apart from these issues, the mobile solution, like drones and robots, can be noisy and allow limited operational life.
As a result, their adoption for human sensing is very limited.
\name takes a hybrid approach where the mobility framework is embedded in the environment, such that it allows for precise relocation without the need to track movement through sensing.

\section{Conclusion}
In conclusion, we propose \name, a gear-based mobility platform for environment-integrated sensors.
\name uses a gear matrix based system to enable 2D sensor movement.
We implement the physical prototype and show the feasibility of the hardware with preliminary characterization.
We also analyze the system performance via simulation for different matrix sizes and numbers of concurrent sensors.

\bibliographystyle{ACM-Reference-Format}
\bibliography{ref}
\balance
\end{document}